# Defect behavior and radiation tolerance of MAB phases (MoAlB and Fe$_2$AlB$_2$) with comparison to MAX phases

Hongliang Zhang[a,§,*], Jun Young Kim[a,§,*], Ranran Su[a], Peter Richardson[c], Jianqi Xi[a], Erich Kisi[c], John O'Connor[c], Liqun Shi[b,*], and Izabela Szlufarska[a,*]

a, Department of Materials Science and Engineering, University of Wisconsin-Madison, WI, USA, 53706
b, Institute of Modern Physics, Fudan University, Shanghai 200433, Peoples' Republic of China
c, School of Engineering, University of Newcastle, University Drive, Callaghan, NSW, 2308, Australia

§ Co-first authors.
* Corresponding authors
Izabela Szlufarska, szlufarska@wisc.edu
Hongliang Zhang, zhlcanes@hotmail.com
Jun Young Kim, junyoungkim729@gmail.com
Liqun Shi, lqshi@fudan.edu.cn

## Abstract

MAB phases are a new class of layered ternary materials that have already shown a number of outstanding properties. Here, we investigate defect evolution and radiation tolerance of two MAB phases, MoAlB and Fe$_2$AlB$_2$, using a combination of experimental characterization and first-principles calculations. We find that Fe$_2$AlB$_2$ is more tolerant to radiation-induced amorphization than MoAlB, both at 150 °C and at 300 °C. The results can be explained by the fact that the Mo Frenkel pair is unstable in MoAlB and as a result, irradiated MoAlB is expected to have a significant concentration of Mo$_{Al}$ antisites, which are difficult to anneal even at 300 °C. We find that the tolerance to radiation-induced amorphization of MAB phases is lower than in MAX phases, but it is comparable to that of SiC. However, MAB phases do not show radiation-induced cracking which is observed in MAX phases under the same irradiation conditions. This study suggests that MAB phases might be a promising class of materials for applications that involve radiation.

**Background and motivation**

Layered ternary materials [1–3] are known to exhibit many outstanding properties, including thermal shock resistance [2,4,5], oxidation resistance [6,7], hardness, strength, and radiation resistance [8,9]. Radiation effects have been studied primarily in MAX phases (M = early transition metal, A = group A element in the periodic table, and X = C or N), which in some cases can maintain their crystalline structures up to a very high radiation dose [8,9]. For instance, $Ti_3(Si/Al)C_2$ was shown to resist radiation-induced amorphization at room temperature up to 25 displacements per atom (dpa). Radiation resistance arises from the ability of the material to efficiently anneal non-equilibrium defects introduced during bombardment of the lattice with neutrons, ions or electrons. Although it is generally agreed upon that the layered structure of MAX phases plays an important role in defect recovery processes, specific mechanisms are still being debated [10–12]. It has been proposed that radiation tolerance of MAX phases could be correlated with the radiation stability of M-X binary, M-A bonding characteristics, A/MX layer ratio, and low antisite formation energy. The question of whether low antisite formation energy is advantageous is particularly interesting. Specifically, while antisites provide an alternative pathway for accommodation of defects, they can also lead to undesirable phase transformation, as is the case for $Ti_3(Si/Al)C_2$, which transforms from hexagonal to a cubic structure during irradiation at room temperature [13]. Moreover, radiation has been shown to lead to surface cracks in many MAX phase materials, such as $Ti_2AlC$ and $Ti_3AlC_2$, even at high temperatures [9]. The aforementioned-radiation-induced phase transformation and microcracking may limit the application of MAX phases in nuclear reactors.

In the current study, we consider another class of layered ternary materials labeled as MAB (M = transition metal, A = Al, B = B) phase. Many MAB phases have been predicted theoretically [14] and some of them have already been synthesized (M = Mo, W, Cr, Mn, Fe, and Ru) [1]. In this work we focus on two MAB materials: MoAlB and $Fe_2AlB_2$. A schematic view of the atomic structure of the two phases is shown in Figure 1. MoAlB is orthorhombic with a space group of *Cmcm*, and $Fe_2AlB_2$ is orthorhombic with a space group of *Cmmm*. Similarly to the layered structure of MAX phases, MAB phases have a metal boride sublattice interleaved by Al layer(s). MoAlB has been shown to have good oxidation resistance thanks to the formation of a protective layer of $Al_2O_3$ [6], and $Fe_2AlB_2$ has been shown to be resistant to cracking [15] and to have a magnetocaloric effect with an ordering temperature of 307 K [16]. Moreover, MoAlB and Fe$_2$AlB$_2$ have high decomposition temperatures of 1708 K [17] and 1500 K [5], respectively.

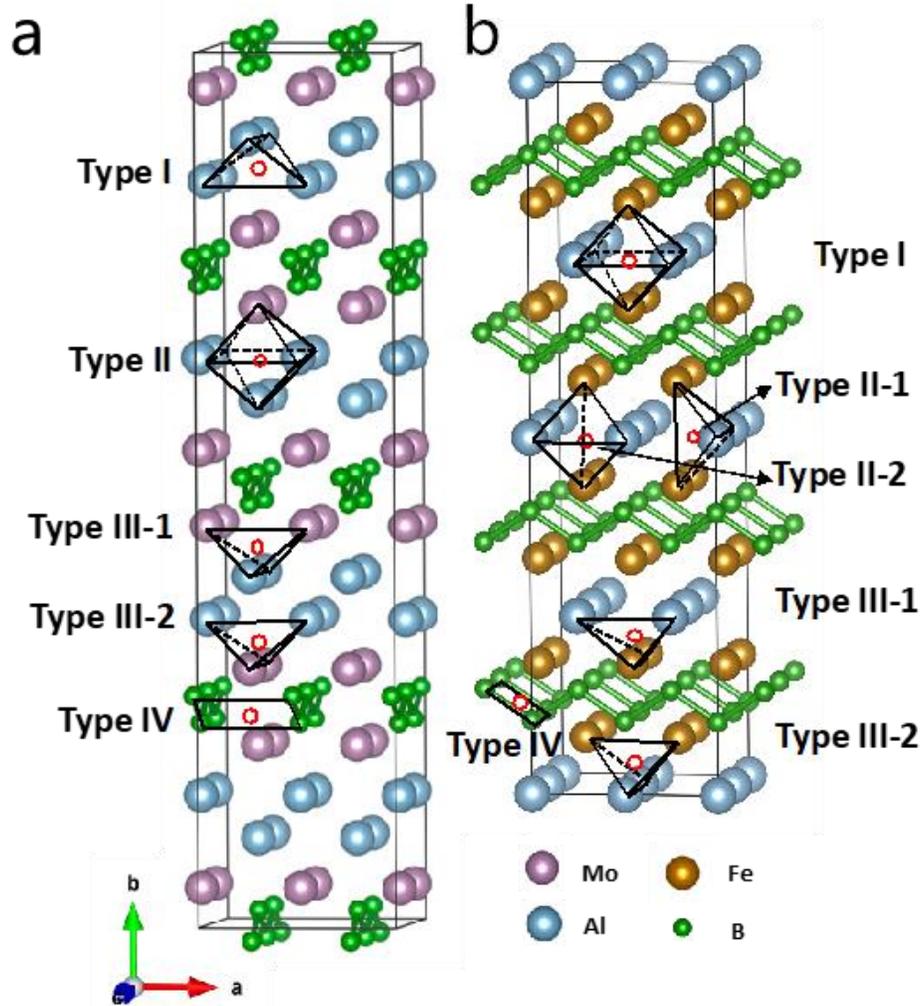

*Figure 1 Schematic view of a supercell (2×2×2) for (a) MoAlB and (b) Fe$_2$AlB$_2$. Different sites for interstitials are indicated as red dots.*

Borides are already considered for applications that involve radiation, such as neutron shielding in both fusion and fission reactors [18,19], and neutron absorbers, e.g., as absorbers of thermal neutrons for long-term, compacted storage of spent nuclear fuel [20]. In order to explore the potential of MAB phases for use in radiation environments, in this study we performed irradiation of the two MAB phases at 150 °C and 300 °C and the radiation effects were analyzed using transmission electron microscopy (TEM). The trends are explained based on calculations of defect formation and migration energies, carried out using density functional theory (DFT). Performance of MAB phases under radiation is compared to that of MAX phases as well as silicon carbide. SiC is known to have good radiation resistance and is already being considered for cladding applications in nuclear reactor applications. The two MAX phases used for comparison are: $Ti_2AlC$ (which contains Al, just like the MAB phases considered in this study) and $Ti_3SiC_2$ (which so far has shown the highest resistance to radiation-induced amorphization among MAX phases).

**Methods**

The materials used in this work were polycrystalline bulk $Fe_2AlB_2$, MoAlB, $Ti_3SiC_2$ and $Ti_2AlC$, prepared by reactive hot press sintering. The polycrystalline 3C-SiC for this research was purchased from Rohm and Haas Company and it had an average grain size of 5 μm [21].

The $Fe_2AlB_2$ sample was made from $Fe_2AlB_2$ powder (<45 μm particle size) synthesized in a tube furnace (MTI GSL-1800X-S60). The powder was placed in a 12.7 mm inner diameter graphite die, lined with a graphite foil, and pressed at a maximum temperature and pressure of 1200 °C and 50 MPa in a hot-press furnace (MTI OTF-1500X-VHP4 containing a mullite tube) under flowing argon to prevent oxidation. The sample was held at the maximum temperature and pressure for 30 min, using a heating and cooling rate of 10 °C /min. The sample was ground using SiC paper to remove graphite from the surface and resulted in a pellet which was 11 mm tall and 12 mm in diameter. The density of the sample determined by Archimedes principle was >95% of theoretical density.

The MoAlB sample was synthesized using MoB powder (<45 μm particle size) and Al powder (>99.7%, <45 μm particle size). The MoB powder was mixed with the Al powder in an atomic ratio of 1:1.3 (MoB:Al). The mixed powder was placed in a 15 mm inner diameter graphite die, lined with graphite foil, and pressed at a maximum temperature and pressure of 1400 °C and 50 MPa in the hot-press furnace (MTI OTF-1500X-VHP4 containing a mullite tube) under flowing argon to prevent oxidation. The sample was held at the maximum temperature for 1 hour, using a heating and cooling rate of 10 °C /min. The sample was ground using SiC paper to remove graphite from the surface and resulted in a pellet which was 95.5% theoretical density, determined by Archimedes principle. The sample was sliced using a diamond blade to produce ~1.5mm thick disks.

Polycrystalline bulk Ti$_2$AlC and Ti$_3$SiC$_2$ were synthesized by the reactive hot-press sintering method [22]. For Ti$_2$AlC, stoichiometric mixtures of Ti + TiAl + C were prepared in argon by hand-grinding a fine powder mixture of Ti (99.5 %, average particle size of 48 µm), Al (99.0 % , average particle size of 48 µm), TiC (99.0 %, average particle size of 5 µm) and C graphite (99.5 %, average particle size of 48 µm) at a molar ratio of 0.5: 1.5: 1: 0.5. The powder mixture was loaded into a cylindrical die, and sintered via hot pressing under flowing argon gas, by heating to 1400 °C at 10 °C/min, and holding at that temperature for 1 h with an applied pressure of 35 MPa. For Ti$_3$SiC$_2$, stoichiometric mixtures of 3Ti + SiC + C were prepared by hand grinding fine Ti (99.9%), SiC (99.9%), and C (graphite, 99.99%) powders under argon, followed by cold pressing in a hardened steel die at 180 MPa. The powders contained ~2 wt.% Al to assist with reactivity. The pressed cylindrical samples were sintered under flowing argon gas by heating to 1600°C at 10°C/min, holding for 4 h, and returning to RT. During sintering, a small amount of Al$_2$O$_3$ was formed in the sample.

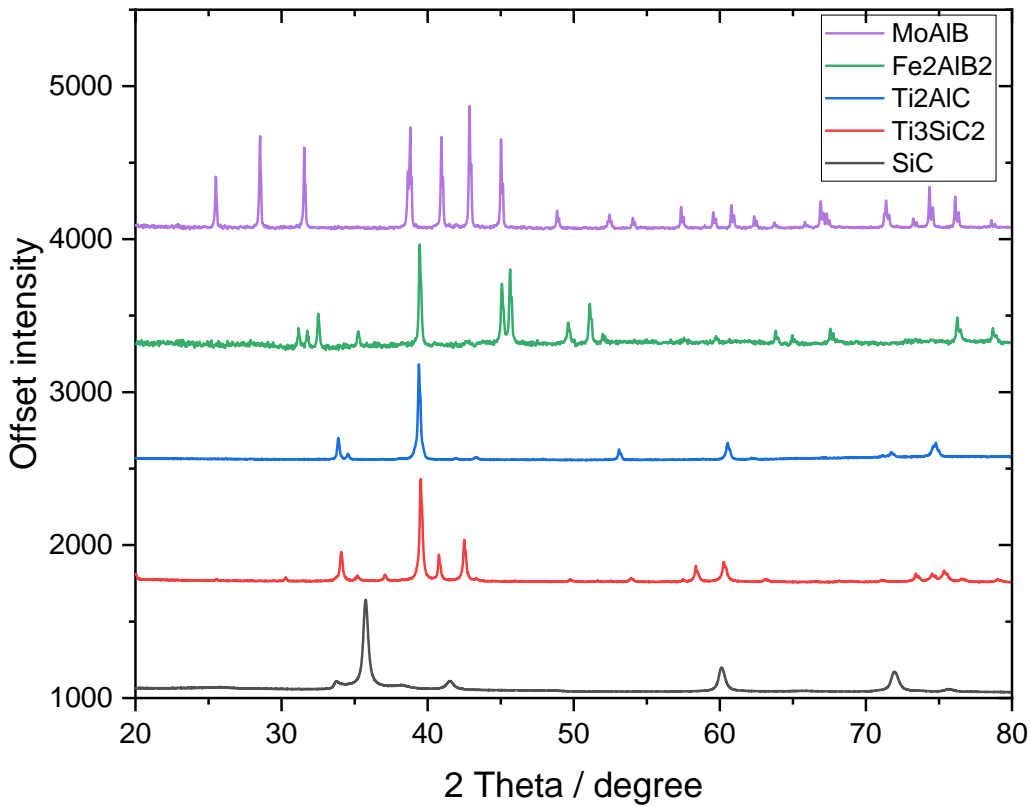

***Figure 2*** *XRD of the unirradiated MoAlB, Fe$_2$AlB$_2$, Ti$_2$AlC, Ti$_3$SiC$_2$, and SiC*

The as-sintered specimens and the purchased CVD SiC were all characterized using XRD (measured by Rigaku D/max 2500, which adopts a Cu Kα source with the wavelength of 0.154 nm with a step size of 0.02°, a time of 1.0 s per step, and a 2-theta range of 20-80 °). As shown in Fig. 2, no impurity phases were found in $Fe_2AlB_2$, MoAlB, $Ti_2AlC$, or SiC, whereas a small amount of $Al_2O_3$ (<2% wt.%) was found in $Ti_3SiC_2$. All specimens were polished using fine metallographic abrasive papers and $Al_2O_3$ suspensions, cleaned by rinsing in ultrasonic baths of acetone and ethanol, and annealed at 600 °C in a vacuum environment of $5 \times 10^{-5}$ Pa for 1 h to release residual stress.

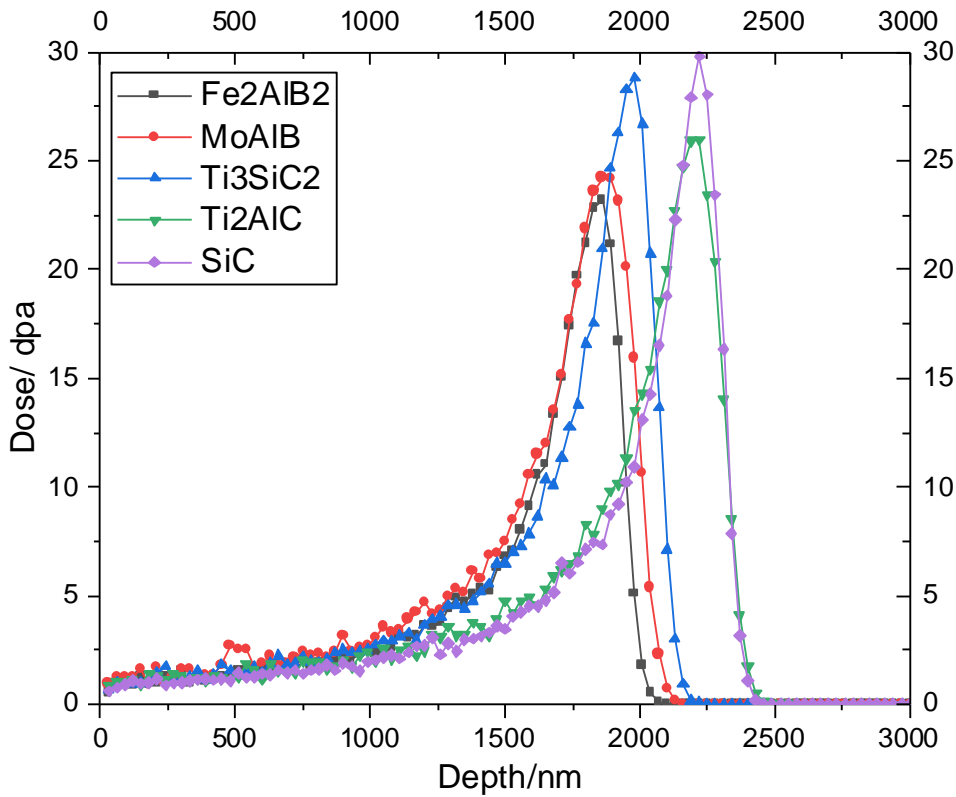

*Figure 3 Irradiation dose and dpa versus depth profiles for $Fe_2AlB_2$, MoAlB, $Ti_2AlC$, $Ti_3SiC_2$ and SiC at a fluence of $1.5 \times 10^{17}$ ions·cm$^{-2}$.*

The final $Fe_2AlB_2$, MoAlB, $Ti_2AlC$, $Ti_3SiC_2$ and CVD SiC bulk samples were irradiated with a 3.15 MeV carbon ion beam incident at 0° to the normal using the tandem accelerator at Ion Beam Lab, Department of Engineering Physics, University of Wisconsin-Madison. Irradiation was performed at 150 °C and at 300 °C. The typical irradiation flux was kept at ~$7.0 \times 10^{11}$ ions·cm$^{-2}$·s$^{-2}$. The irradiation fluence delivered to the samples was $1.5 \times 10^{17}$ ions·cm$^{-2}$ for the high dose case and 7.5 ×10$^{16}$ ions·cm$^{-2}$ for the low dose case. The background pressure during irradiation was < 5×10$^{-4}$ Pa. The total damage, measured in dpa, was simulated using SRIM-2013 [23]. The displacement energies are 25 eV, 25 eV, and 28 eV, respectively, for Fe, Al, and B in Fe$_2$AlB$_2$; 25 eV, 25 eV, and 28 eV, respectively for Mo, Al, and B in MoAlB; 25 eV, 15 eV and 28 eV, respectively, for Ti, Al, and C in Ti$_2$AlC; 25 eV, 15 eV, and 28 eV, respectively, for Ti, Si, and C in Ti$_3$SiC$_2$; and 15 eV for both Si and C in SiC. The damage level obtained from the SRIM-2013 simulation was estimated to be 1.0 dpa at the surface, rising to ~23 dpa for Fe$_2$AlB$_2$ and MoAlB at a depth of ~1800 nm, ~26 dpa for Ti$_2$AlC, ~28 dpa for Ti$_3$SiC$_2$ and 30 dpa for SiC at a depth of 2200 nm (see Fig. 3). The experimental ranges for the irradiation obtained from the TEM images are 2.2 µm for Fe$_2$AlB$_2$ and MoAlB, 2.4 µm for Ti$_3$SiC$_2$ and 2.6 µm for Ti$_2$AlC and CVD SiC.

Scanning electron microscopy (SEM) images and samples for TEM analysis were obtained using standard lift-out techniques by a FEI Helios PFIB G4 FIB/FESEM Focused Ion Beam (FIB) instrument in the Materials Science Center at the University of Wisconsin-Madison [24]. To protect the sample surface from damage during FIB preparation, a 3.0 µm Pt protective layer was deposited on the surface of the indented region by two steps: (i) 2 kV electron beam (low energy) was used to deposit a 1.0 µm Pt layer to avoid damage from high-energy ions deposition; (ii) a 12 kV ion beam was used for the deposition of another 2.0 µm Pt layer. The thinning process was accelerated by using a high-energy ion beam (30 kV) at the beginning and a low-energy ion beam (2 kV) at the end to carefully remove the amorphous area generated in the former stage. A FEI Tecnai F30 with field emission gun (FEG) TEM and high resolution TEM were used to analyze the damage and microstructural changes before and after irradiation.

DFT calculations were performed using the Vienna Ab-initio Simulation Package (VASP) [25] with the projector augmented wave (PAW) [26] and the generalized gradient approximation (GGA) by Perdew, Burke and Ernzerhof (PBE) [27]. The plane-wave cutoff energy of 400 eV and Monkhorst-Pack k-point mesh of $8 \times 2 \times 8$ were set with an energy tolerance of 0.7 meV/atoms [28]. The lattice constants of the two MAB phases were calculated as: MoAlB ($a = 3.215$ Å, $b = 14.035$ Å, $c = 3.109$ Å) and Fe$_2$AlB$_2$ ($a = 2.913$ Å, $b = 11.007$ Å, $c = 2.861$ Å), both in good agreement with the experimentally determined lattice constants [1]. Total energies of perfect structures as well as those containing either a vacancy or an antisite were calculated in $2 \times 2 \times 2$ supercells. Interstitial calculations required larger supercells, which were determined based on results of a convergence test. The reported values were determined using the following supercells: $4 \times 2 \times 4$ ($Mo_I$ and $Al_I$ in MoAlB and $B_I$ in $Fe_2AlB_2$), $4 \times 1 \times 4$ ($B_I$ in MoAlB), $6 \times 1 \times 6$ ($Fe_I$ in $Fe_2AlB_2$), and $6 \times 2 \times 6$ ($Al_I$ in $Fe_2AlB_2$), where we use the Kröger-Vink notation to label point defects.

The defect formation energy ($E_f$) was calculated using the following equation,

$$E_f = E_\text{defective} - E_\text{perfect} + n_i \mu_i \quad (i = \text{Mo, Fe, Al or B}) \quad \text{Eq. 1}$$

where $E_\text{defective}$, $E_\text{perfect}$, $n_d$, and $\mu_d$ represent the total energy of a defective supercell, the total energy of a perfect supercell, the number of defective atoms, and the chemical potential of a defective atom, respectively. In addition to the formation energies, migration energies of vacancies and interstitials were calculated. Reaction energies were calculated to consider possible reactions between defects. Migration energies and reaction energy barriers were calculated using the climbing image nudged elastic band method [29]. Finally, interlayer binding energies between M and A layers were calculated by subtracting the energy of a perfect unit cell from the energy of a unit cell where a gap of 1 nm is inserted between a M and an A layer, and by dividing the calculated value by the number of surface atoms.

**Results and Discussion**

Damage introduced during irradiation was analyzed using cross-sectional TEM and HRTEM images shown in Fig. 4. The top and bottom rows correspond to MoAlB and $Fe_2AlB_2$, respectively. Left, middle, and right column correspond to irradiation of $7.5 \times 10^{16}$ ions·cm$^{-2}$ at 150 °C (low dose, 150 °C), $1.5 \times 10^{17}$ ions·cm$^{-2}$ at 150 °C (high dose, 150 °C), and $1.5 \times 10^{17}$ ions·cm$^{-2}$ at 300 °C, respectively. The images show that MoAlB irradiated at 150 °C (both low and high dose) becomes entirely amorphous (Fig. 4a), as the selected area electron diffraction (SAED) pattern in Fig. 4a shows diffuse rings with no indication of diffraction spots anywhere in the irradiated region and no contrast in dark field imaging, which is typical of amorphous material. Amorphization of the structure has been further confirmed by the HRTEM shown in Fig. 4d. Irradiation of MoAlB at 300 °C also led to complete amorphization of the structure (see Figs. 4c and 4f).

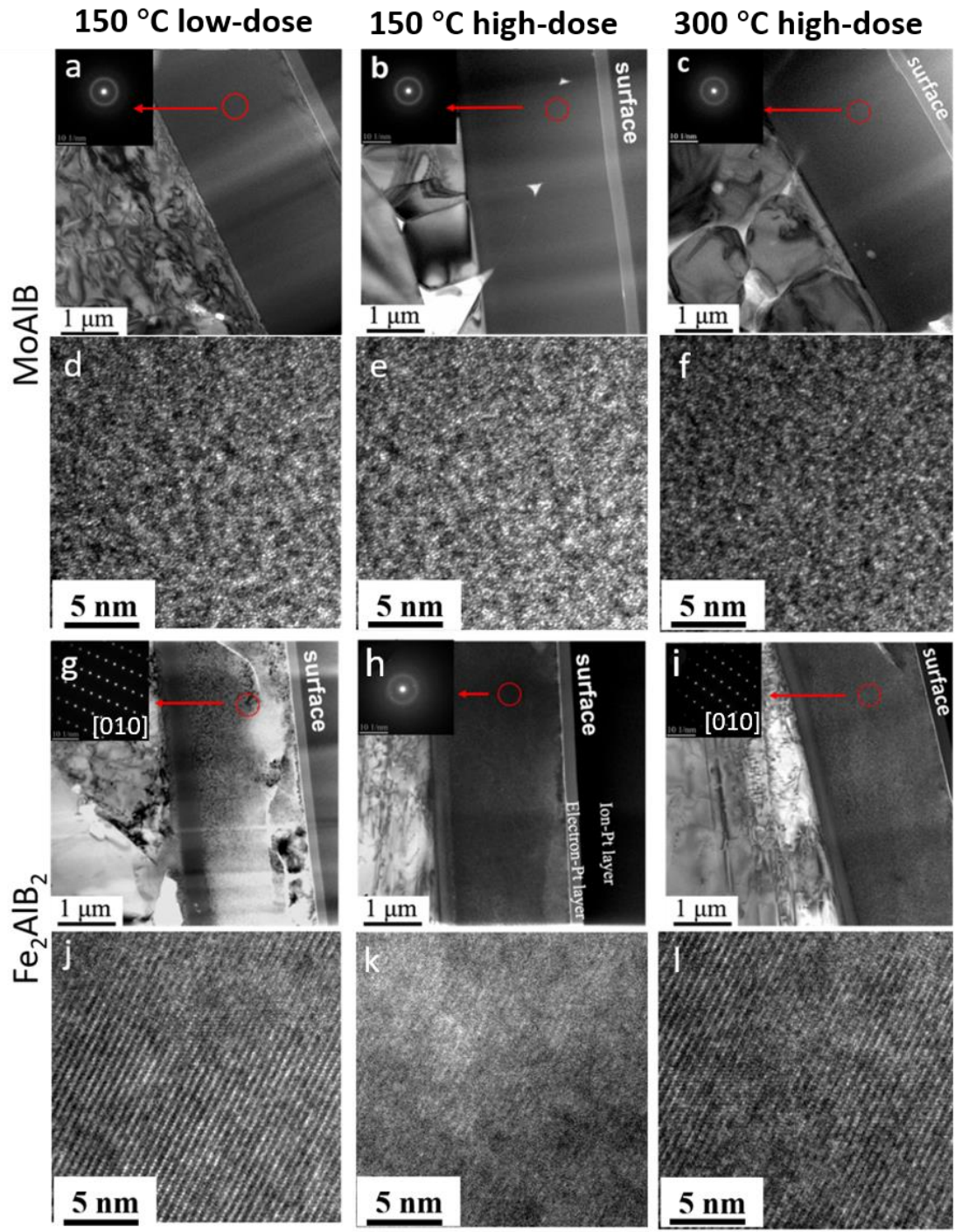

*Figure 4* TEM and HRTEM images of MoAlB (top row) and Fe$_2$AlB$_2$ (bottom row), irradiated at 7.5 ×10$^{16}$ ions·cm$^{-2}$ at 150 °C (left column), at 1.5 ×10$^{17}$ ions·cm$^{-2}$ at 150 °C (middle column), and at 1.5 ×10$^{17}$ ions·cm$^{-2}$ at 300 °C (right column). The light-colored thin band on the top of the surface is the Pt protective layer deposited by electron deposition followed by the thicker, dark colored Pt protective layer deposited by ion deposition during FIB.

Fe$_2$AlB$_2$ showed a much better resistance to radiation-induced amorphization under the same irradiation conditions. Specifically, after the low dose 150 °C irradiation Fe$_2$AlB$_2$ remained crystalline as evidenced by the SAED patterns shown in the inset of Fig. 4g and the HRTEM image shown in Fig. 4j. The irradiation produced many black-spot defects (corresponding to defect clusters), and the density of black spot defects is relatively high in the region that experienced a higher dose in Fig. 4g. However, there is no evidence of amorphization in most parts of the sample except for the damage peak region where there is an amorphous band with width of ~0.2 μm, corresponding to a damage dose of ~11 dpa. As the irradiation fluence increased to $1.5\times10^{17}$ ions·cm$^{-2}$ at 150 °C, most of the Fe$_2$AlB$_2$ became significantly damaged but there were still some crystalline structures remaining as indicated by the HRTEM in Fig. 4k and by the weak diffraction spots in the SAED pattern in the inset of Fig. 4h. There is an obvious amorphization band with a width of 0.3 μm around the damage peak area corresponding to the peak damage dose of ~23 dpa. After irradiation at 300 °C, Fe$_2$AlB$_2$ remained crystalline, as evidenced by both, the HRTEM image in Fig. 4l (which shows a highly ordered structure with some black spot defects) and the SAED pattern in the Fig. 4f inset (which shows clear diffraction spots without rings). This result indicates that the threshold dose for amorphization is larger at 300 °C than at 150 °C. The width of the amorphous band is ~0.2 μm and the edge of the band in the near surface direction corresponds a damage dose of ~10 dpa. Combining our experimental results and the SRIM calculations, for 3.15 MeV carbon ions, the fluence to amorphize the flat region of Fe$_2$AlB$_2$ can be roughly estimated as $2.0\times10^{17}$ ions·cm$^{-2}$ at 150 °C, and $3.5\times10^{17}$ ions·cm$^{-2}$ at 300 °C. The dose to amorphization for Fe$_2$AlB$_2$ is about 10 dpa at 150 °C and 16 dpa at 300 °C. Since MoAlB became amorphous at all fluences, it was difficult to obtain an accurate estimate of the fluence to amorphization. The fluence to amorphize the flat region of MoAlB is less than $7.5\times10^{16}$ ions·cm$^{-2}$.

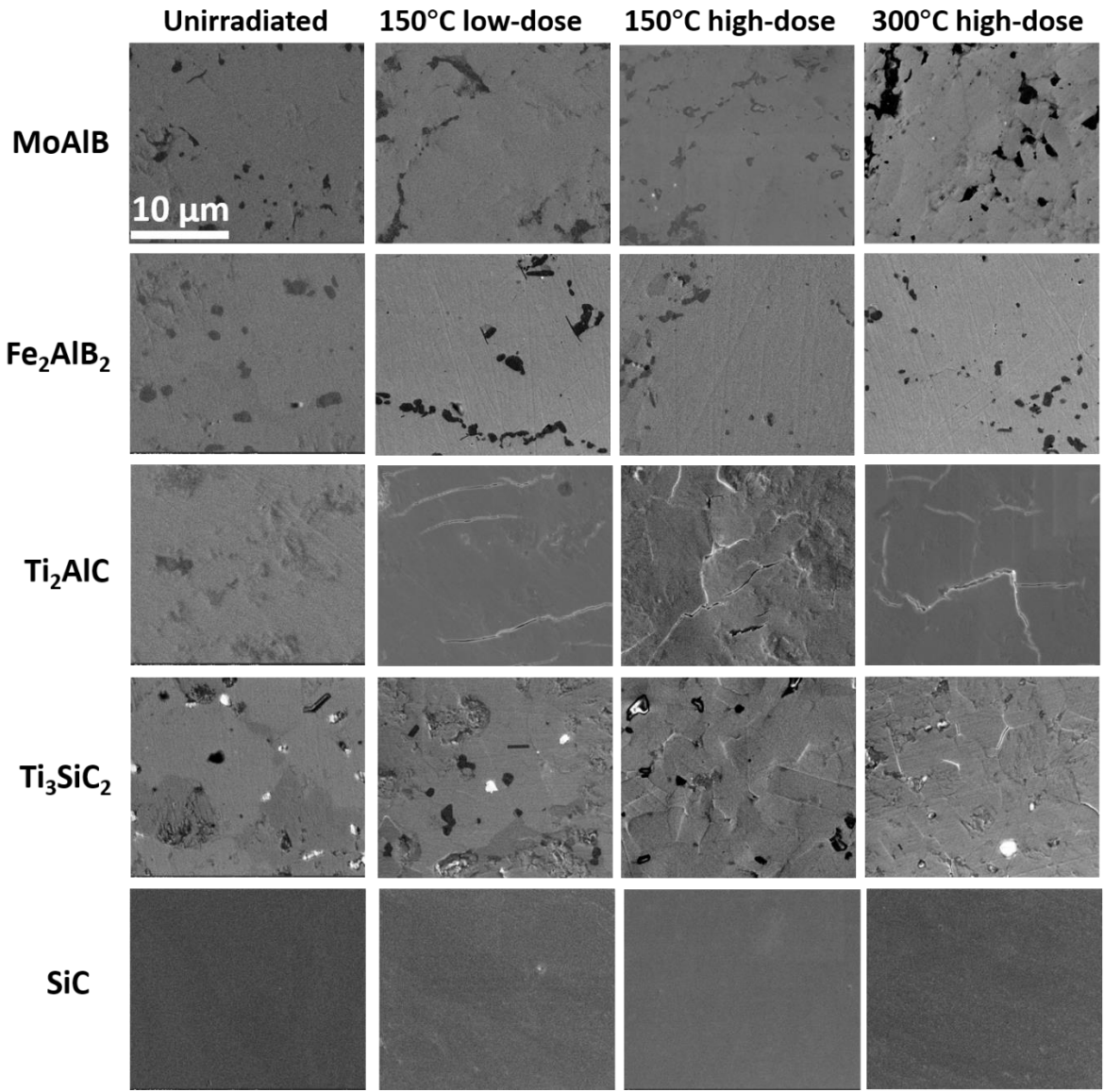

*Figure 5* *SEM images of the MoAlB, Fe₂AlB₂, Ti₂AlC and Ti₃SiC₂ MAX phase, and CVD SiC unirradiated and irradiated at 150 °C low dose, 150 °C high dose and 300 °C high dose. The scale bars in all SEM images are identical.*

The radiation-induced cracks in $Ti_2AlC$ and $Ti_3SiC_2$ are believed to be caused by the anisotropic swelling, i.e., the lattice swelling along the *c* axis and contraction along the *a* axis [30–33]. To determine if there is radiation-induced anisotropic swelling in $Fe_2AlB_2$ (MoAlB was fully amorphous after irradiation, so we did not analyze it), GIXRD spectra at an incident angle of 1.0 degree were collected and they are shown in Fig. 6. Unlike the GIXRD results of the irradiated MAX phase (e.g., Ref. [30]), where a significant shift of peaks has been reported, no obvious shift of peaks' positions was found by us in the spectra of the irradiated $Fe_2AlB_2$, indicating the change of the lattice parameter was very small. The refinement results of the GIXRD also showed that the irradiation-induced lattice-parameter (LP) change in $Fe_2AlB_2$ was minimal ($a$-LP increased by 1.2%, $b$-LP decreased by 0.9 %, and $c$-LP decreased by 0.9%) as compared to that of the MAX phase (a-LP decreased by ~1.0%, and c-LP increased by ~4% [30]). For the 150 °C-high dose irradiated $Fe_2AlB_2$ sample, $a$-LP changed from 2.913 to 2.949 Å, with an increase of 1.2%, whereas $b$-LP changed from 11.003 to 10.903 Å, a slight decrease of 0.9%. The $c$-LP decreased from 2.861 Å to 2.836 Å, which is again a slight decrease of 0.9%. The LP changes of the 150 °C -low dose and 300 °C high dose irradiated samples are even smaller (less than 0.7% for the increase of $a$-LP, and <0.3% for the decrease of $b$-LP and $c$-LP). The results show that although there are slightly anisotropic changes in the LP of $Fe_2AlB_2$ in $Fe_2AlB_2$, these changes are much smaller than those reported in $Ti_3SiC_2$ and $Ti_2AlC$. Moreover, for the MAX phase, $a$-LP decreased, and the $c$-LP increased after the irradiation, whereas for $Fe_2AlB_2$ MAB phase, $a$-LP increased, $b$ and $c$-LP decreased after irradiation. The lower overall as well as the lower swelling anisotropy in $Fe_2AlB_2$ could be the reason for the lack of radiation-induced cracks in $Fe_2AlB_2$. SiC was found to be free of irradiation-induced cracks for all doses at and temperatures considered in this study.

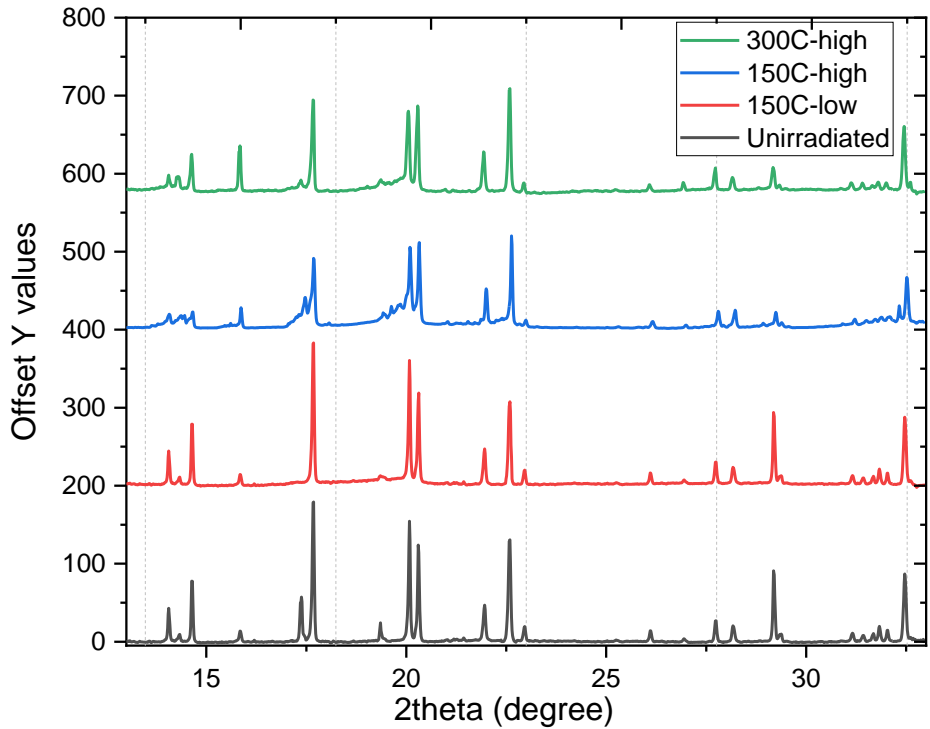

*Figure 6 GIXRD spectra at an incident angle of 1.0° for $Fe_2AlB_2$ unirradiated, irradiated at $7.5 \times 10^{16}$ ions·cm$^{-2}$ at 150 °C (150 °C -low), at $1.5 \times 10^{17}$ ions·cm$^{-2}$ at 150 °C (150 °C -high), and at $1.5 \times 10^{17}$ ions·cm$^{-2}$ at 300 °C (300C-high).*

We have not found any radiation-induced phase transformation in $Fe_2AlB_2$. Such phase transformation has been reported in irradiated MAX phases [13,34]. Specifically, GIXRD spectra collected for $Fe_2AlB_2$ (see Fig. 6) show that no new peak was generated after irradiation. SAED results from [010] direction show there are no diffraction spots from other crystal structures, and the HRTEM results from the same direction show the structure is the same except for the difference in the defect densities and the level of disorder in the samples irradiated at different doses. All these results support the conclusion that there is no irradiation-induced phase transformation in $Fe_2AlB_2$.

The experimentally observed trends in the tolerance to radiation-induced amorphization of the two MAB phases can be rationalized based on first-principles calculations. First of all, to calculate the defect formation energy using Eq. 1, we determined chemical potentials of constituent elements (Mo, Fe, Al, and B) in the two MAB phases, as shown in Fig. 7. The yellow lines are the precipitation lines of the binary phases, and the highlighted areas indicate the chemical potential ranges where MAB phases can be formed without precipitating other phases [see S.1. for detailed calculations and values]. The line EF was chosen first to reflect the precipitated binaries $Al_8Mo_3$ and $Al_{13}Fe_4$, which are the predominant precipitates found in experiments [4,35], and then the line XY was selected in order to compare the two MAB phases at the same reference points; the two lines share the same chemical potential of Al $\mu(Al)$.

While each of the vacancy and antisite defects has only one possible site in MoAlB and $Fe_2AlB_2$, interstitials can potentially occupy several different positions. The potential interstitial sites investigated in our study are depicted in Fig. 1. The positions we explored for MoAlB are Type I (a center of a tetrahedron with four Al neighbors), Type II (a center of an octahedron with five Al and one Mo neighbors), Type III (a center of a tetrahedron with two Al and two Mo neighbors, having two cases depending on the orientation of the tetrahedron), and Type IV (a center of a rectangle with four B neighbors). These interstitial sites were tested by placing an atom (Mo, Al, or B) near the center of the potential interstitial site with a slight (0.15 Å) displacement from the symmetric point[36]. Our calculations show that an Al interstitial forms in Type I only. When an Al atom is placed in Type II or Type III positions, the configuration is unstable and spontaneously relaxes into Type I. $Al_I$ of Type IV was also found to be unstable. Specifically, in this case an Al interstitial kicked out a Mo atom from its lattice site and formed an antisite $Al_{Mo}$.

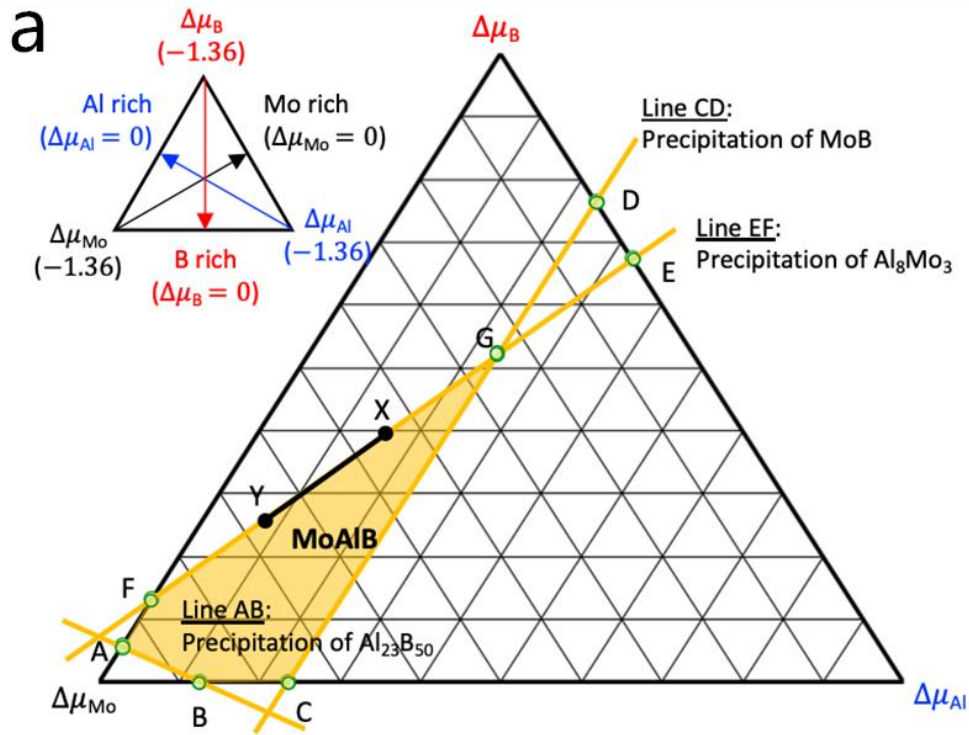
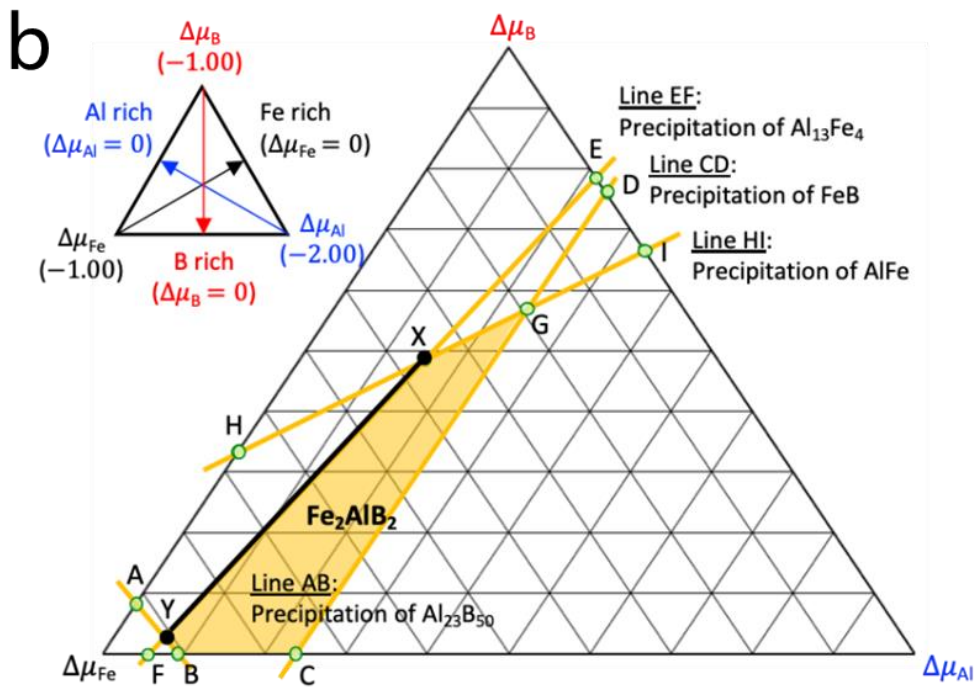

*Figure 7* Chemical potential map of (a) MoAlB and (b) Fe$_2$AlB$_2$ with the highlighted area indicating the chemical potential ranges where MAB phases can be formed without precipitating other phases.

Simultaneously, the displaced Mo atom kicked out another nearby Al atom from its lattice site, which finally resulted in the second Al atom taking an interstitial position of Type I. For $B_I$ in MoAlB, the most stable position (lowest formation energy) is also of Type I, followed by Type II ($E_f$ higher by 1.29 eV than Type I), Type III-1 ($E_f$ higher by 1.35 eV than Type I), and Type IV ($E_f$ higher by 2.26 eV). $B_I$ is not stable for Type III-2, and instead it relaxes into Type I. Interestingly, Mo has no sites for interstitial. Testing of Type I-IV revealed that a Mo interstitial kicks out an Al atom and becomes an antisite defect $Mo_{Al}$, making the kicked-out Al atom $Al_I$ of Type I.

Similar types of interstitials were considered for $Fe_2AlB_2$ (see Fig. 1). These were Type I (a center of an octahedron with four Al and two Fe neighbors), Type II (a center of a tetrahedron with two Al and two Mo neighbors, having two cases depending on the orientation of the tetrahedron), Type III (a center of a tetrahedron with two Al and two Fe neighbors, having two cases depending on the orientation of the tetrahedron), and Type IV (a center of rectangle with four B neighbors). We found that Al does not form any of these types of interstitial defects. Specifically, when an Al atom is placed on Type I-III positions, it spontaneously relaxes into the Al layer and forms a dumbbell-like interstitial with another Al atom. An Al atom placed on a Type IV site kicks out a Fe atom and becomes an antisite defect $Al_{Fe}$. The displaced Fe atom forms an interstitial $Fe_I$ of Type II-1. For $B_I$ in $Fe_2AlB_2$, the most stable position is Type I, followed by Type III-2 ($E_f$ higher by 1.22 eV), Type IV ($E_f$ higher by 1.58 eV), and Type III-1 ($E_f$ higher by 8.20 eV). Placing B in Type II revealed that $B_I$ is not stable on these sites, instead it relaxes to the configuration of Type I. Lastly, $Fe_I$ forms only on Type II; the most stable is the Type II-1 configuration, followed by Type II-2 ($E_f$ higher by 0.63 eV than Type II-1). Fe placed on any other interstitial site relaxes into Type II-1.

In summary, the most stable interstitials for MoAlB and $Fe_2AlB_2$ are all located in the Al layer, as described in Supporting Information 2, which has been observed in MAX phases as well [12]. Note that $Mo_I$ does not form because $Mo_I$ is unstable and instead forms $Mo_{Al}$ and $Al_I$. The defect configuration found in this study implies that most Frenkel Pairs (FPs) likely form in Al layers and those Al layers play a crucial role in accommodating defects, similarly to what has been reported MAX phases. Therefore, our study focuses on interstitial defects only in the Al layer.

**Table 1** *Formation energies of point defects in MoAlB and Fe$_2$AlB$_2$. Energies are referenced to chemical potentials along the XY line in Fig. 5.*

| M = Mo, Fe | Formation energy (eV) | |
| --- | --- | --- |
| | MoAlB | Fe$_2$AlB$_2$ |
| V$_M$ | 1.94-2.26 | 0.55-1.05 |
| V$_{Al}$ | 1.00-1.12 | 2.11-2.23 |
| V$_B$ | 0.56-0.76 | 0.07-0.51 |
| M$_I$ | Mo$_I$ → Mo$_{Al}$ + Al$_I$ | 4.86-5.36 |
| Al$_I$ | 6.42-6.54 | 5.59-5.71 |
| B$_I$ | 3.06-3.26 | 2.26-2.70 |
| M$_{Al}$ | 2.13-2.57 | 0.85-1.47 |
| M$_B$ | 6.97-7.49 | 3.47-4.41 |
| Al$_M$ | 1.77-2.21 | 1.34-1.96 |
| Al$_B$ | 2.44-2.52 | 2.84-3.16 |
| B$_M$ | 3.92-4.44 | 2.78-3.72 |
| B$_{Al}$ | 1.66-1.74 | 2.66-2.98 |
| M$_{FP}$ | unstable | 2.96 |
| Al$_{FP}$ | 3.77 | 3.91 |
| B$_{FP}$ | 1.91 | 1.39 |

**Table 2** *Migration energies of vacancies and interstitials in MoAlB and Fe$_2$AlB$_2$.*

| M = Mo, Fe | Migration energy (eV) | | | | | |
| --- | --- | --- | --- | --- | --- | --- |
| | V$_M$ | V$_{Al}$ | V$_B$ | M$_I$ | Al$_I$ | B$_I$ |
| MoAlB | 5.27 | 0.46 | 0.68 | X | 0.37 | 1.40 |
| Fe$_2$AlB$_2$ | 3.34 | 1.17 | 0.50 | 1.07 | 0.25 | 0.92 |

In Table 1, we report the formation energies calculated using Eq. 1 on the line XY from Fig. 7. The formation energies of vacancies, interstitials, and antisites vary depending on the chemical potentials, whereas that of FPs is independent of the chemical potentials. As mentioned earlier, Mo$_I$ does not form in MoAlB, but instead it relaxes to Mo$_{Al}$ and Al$_I$, thus Mo$_{FP}$ is labeled as "unstable."

Table II shows the calculated migration energies. We report only migration energies for vacancies and interstitials, because direct migration of an antisite defect would involve an exchange with a neighboring atom on the lattice, which is energetically prohibitive. The tabulated migration energies are the lowest energies among the possible migration paths. $V_{Al}$ in MoAlB has the lowest migration energy (0.46 eV) when migrating into the nearest diagonal site within an Al layer, whereas the migration along the a-axis and c-axis (see Fig. 1) have the energies of 3.45 eV and 2.98 eV, respectively. $V_B$ migrating along the c-axis has a migration energy of 0.68 eV. The migration of $V_{Mo}$ is very unlikely; the lowest energy (5.27 eV) is obtained when $V_{Mo}$ moves along the c-axis in the same Mo plane, whereas migration along the a-axis in the same Mo plane and into another Mo plane through B layers have energies of 5.54 eV and 6.67 eV, respectively. The migration energy of $V_{Mo}$ can decrease when there are nearby $Mo_{Al}$ that exert a repulsive force on $V_{Mo}$ (hence, attractive force to a migrating Mo atom). When $Mo_{Al}$ is located at the nearest site to $V_{Mo}$, the migration energy of $V_{Mo}$ is calculated to be 4.84 eV, while that of $V_{Mo}$ with $Mo_{Al}$ located at the 2nd nearest site is 5.03 eV.

In the case of interstitials, we only consider the migration of the most stable interstitial into symmetry- equivalent positions, namely, these are migrations within the Al layer. $Al_I$ in MoAlB has three migration paths: along the a-axis, the c-axis, and a diagonal line within the Al layer, with the energy barriers of 0.37 eV, 1.25 eV, and 2.43 eV, respectively. Next, $B_I$ in MoAlB has the same three migration paths: along the a-axis, the c-axis, and a diagonal line within the Al layer, with the energy barriers of 1.40 eV, 1.65 eV, and 2.22 eV, respectively.

As for defect migration in $Fe_2AlB_2$, $V_{Al}$ has two migration paths: along the a-axis (1.17 eV) and along the c-axis (1.51 eV) in the same Al plane. $V_B$ can migrate along the a-axis with the activation energy of 0.50 eV, and $V_{Fe}$ has three possible migration paths: along the a-axis (4.42 eV), the c-axis (5.02 eV), and into another Fe plane through B layers (3.34 eV). $Al_I$ can migrate by transitioning from one dumbbell position to another: along the c-axis (0.25 eV) and along the a-axis (0.36 eV). $B_I$ has two paths of migrating into another octahedral site: along the a-axis (1.20 eV) and along the c-axis (with the migration energy of 0.92 eV). Lastly, $Fe_I$ has two paths of migration into another tetrahedral site: along the a-axis and the c-axis, with the corresponding migration energies of 1.42 eV and 1.07 eV.

Previous theoretical studies have rationalized the radiation tolerance of ternary MAX phases based on such parameters as the radiation stability of the corresponding M-X binaries (for instance, TiC for $Ti_3SiC_2$), M-A bonding characteristics (the weaker the bond, the better the radiation resistance), the ratio of the number of A and MX layers (the higher the ratio, the better the radiation resistance), and the formation energy of the $M_A$-$A_M$ pair (the lower the formation energy of the antisite, the better the radiation resistance) [10–12]. Similar analysis in the MAB phases is not possible at this point, because of the limited research to date on radiation effects in MAB phase materials. It is, however, still instructive to ask if the criteria proposed for MAX phases are consistent with our observation that $Fe_2AlB_2$ has shown a better radiation resistance to amorphization than MoAlB. First of all, rationalizing radiation resistance of ternary MAB phases based on radiation studies of the corresponding M-B binaries cannot be tested here because studies of radiation-induced amorphization of MoB or FeB have not been reported in literature, except for $Fe_3B$. $Fe_3B$ irradiated at 112 °C showed partial amorphization at the fluence of $10^{19}$ ions/cm$^2$ and full amorphization at the fluence of $2 \times 10^{19}$ ions/cm$^2$ [37]. Although $Mo_2B_5$ was studied with the irradiation of $10^{19}$ ions/cm$^2$ [38], the focus of the study was on radiation-induced swelling and fracturing, and not on radiation-induced amorphization. Secondly, we tested the hypothesis related to the M-A bonding energy, calculated in this study for MoAlB and $Fe_2AlB_2$ to be 1.67 eV and 1.12 eV, respectively. Thus, the lower bonding energy of $Fe_2AlB_2$ possibly contributes towards the tolerance to radiation-induced amorphization. However, it is still difficult to conclude whether the M-A bonding energy is indeed responsible for the observed trend in radiation resistance. For instance, $Ti_3SiC_2$ is known to be more tolerant to radiation-induced amorphization than $Ti_2AlN$, but the M-A bonding energy of the latter is weaker [12]. Next, the density of A layers is 1/3 for MoAlB and 1/5 for $Fe_2AlB_2$, so the higher density of the A layers in MoAlB is consistent with the hypothesis that the fraction of A layers correlates with the tolerance to radiation induced amorphization. As for the last criterion, the lower $M_A$-$A_M$ pair formation energy of $Fe_2AlB_2$ (2.81 eV) than MoAlB (4.34 eV) could potentially contribute towards the tolerance to radiation-induced amorphization, but it cannot be solely used to determine the tolerance trends because of counter examples. For instance, the $M_A$-$A_M$ pair formation energies in $Ti_3SiC_2$ and $Cr_2AlC$ are 3.52 eV and 2.40 eV, respectively, but the former is known to be more tolerant to radiation-induced amorphization.

To summarize, some of the criteria proposed for radiation resistance of MAX phases are consistent with results of our experimental studies on MAB phases (i.e., formation energies of the antisite pair and the bond characteristics), some are not (i.e., the ratio of the number of A and MB layers), and in some cases there is no data (i.e., there are no consistent studies of the corresponding binaries). However, more extensive studies on multiple MAB phase systems will be needed to determine whether there exist simple correlations between fundamental defect properties and the radiation resistance across the different MAB phases. There are also some key differences between MAX and MAB phases. For instance, if other MAB phases do not undergo phase transformation driven by antisite defects (consistently with what we found for $Fe_2AlB_2$), then perhaps formation energies of antisite defects are not the determining factor in radiation resistance to amorphization of these materials.

To rationalize observations from our experiments on the MAB phases and to understand how radiation-induced damage can be annealed, we have analyzed defect energetics in more detail, including defect migration and reaction energies.

*Table 3* *Reaction energies and energy barriers for reactions between point defects in MoAlB and $Fe_2AlB_2$. "Diff" means a diffusion-limited reaction and the number in parenthesis is the lower of the migration energies of the reactant defects. Negative reaction energy means that the reaction is energetically favorable, and "unstable" means that those reactions cannot occur because they involved $Mo_I$, which is unstable and spontaneously transforms into $Mo_{Al}$ and $Al_I$ (see Table 1).*

| # | Reaction (M = Mo, Fe) | Reaction energy (eV) | | Reaction energy barrier (eV) | |
|---|---|---|---|---|---|
| | | MoAlB | $Fe_2AlB_2$ | MoAlB | $Fe_2AlB_2$ |
| 1 | $Al_B + V_{Al} \rightarrow V_B + Al_{Al}$ | −2.89 | −4.88 | Diff (0.46) | Diff (1.17) |
| 2 | $Al_M + V_{Al} \rightarrow V_M + Al_{Al}$ | −0.95 | −3.02 | 0.19 | Diff (1.17) |
| 3 | $B_{Al} + V_B \rightarrow V_{Al} + B_B$ | −1.30 | −0.94 | Diff (0.68) | 0.12 |
| 4 | $B_M + V_B \rightarrow V_M + B_B$ | −2.74 | −2.74 | Diff (0.68) | 0.07 |
| 5 | $M_{Al} + V_M \rightarrow V_{Al} + M_M$ | −3.39 | +0.21 | 1.36 | 1.81 |
| 6 | $M_B + V_M \rightarrow V_B + M_M$ | −8.67 | −4.45 | Diff (5.27) | Diff (3.34) |
| 7 | $Al_I + V_{Al} \rightarrow Al_{Al}$ | −6.82 | −7.82 | Diff (0.37) | Diff (0.25) |
| 8 | $M_I + V_M \rightarrow M_M$ | unstable | −5.91 | unstable | Diff (1.07) |
| 9 | $B_I + V_B \rightarrow B_B$ | −3.82 | −2.77 | 1.33 | 1.02 |
| 10 | $M_I + V_{Al} \rightarrow M_{Al}$ | unstable | −6.12 | unstable | Diff (1.07) |
| 11 | $B_I + V_{Al} \rightarrow B_{Al}$ | −2.52 | −1.83 | Diff (0.46) | Diff (0.92) |
| 12 | $Al_I + M_{Al} \rightarrow M_I + Al_{Al}$ | unstable | −1.71 | unstable | Diff (0.25) |

| 13 | $Al_I + B_{Al}$ → $B_I + Al_{Al}$ | −4.30 | −6.00 | Diff (0.37) | Diff (0.25) |
| 14 | $M_B + V_{Al}$ → $M_{Al} + V_B$ | −5.39 | −4.60 | Diff (0.46) | Diff (1.17) |

In order to determine how different defect recovery processes can lead to radiation resistance of MAB phases, ideally one would build a detailed rate theory model [39–41]. However, development of such a model is beyond the scope of this project. Instead, we consider specific reactions between point defects to determine if radiation resistance of the two MAB phases can be explained by the presence of defects that are difficult to anneal out. Possible reactions between different point defects are listed in Table 3 together with the reaction energy and the reaction energy barrier. "Diff" denotes a diffusion-limited reaction, in which reactions occur spontaneously when the defects are near each other. Note that the reactions involving $Mo_I$ do not occur because $Mo_I$ is unstable and spontaneously transforms into $Mo_{Al}$ and $Al_I$ (see Table 1).

On the basis of Tables 1–3, we identified defect species that are easily annealed by specific reactions and others that cannot be removed and likely remain in the lattice at the temperatures of our experiments, i.e., 150 °C and 300 °C. By calculating the hopping rate, we define the energy range in which the reactions can occur rapidly at a given temperature and therefore the range in which defects can be removed.

In order to determine whether a defect is able to be annealed (e.g., by reacting with other defects), we assume that it has to move at least 1 nm over a reasonable time period (here assumed to be 100 s). Taking this criterion and the pre-exponential factor to be $10^{13}$ s$^{-1}$ one can estimate that defects with the energy barrier of 1.17 eV (82 s) can be annealed. We will assume that defects with migration energy or reaction energy barrier higher than 1.33 eV (104 min) will take much longer to anneal – 76 times longer than reactions with barriers of 1.17 eV. Note, that there is no reaction whose migration or reaction energy barrier is in the range between 1.17 and 1.33 eV. At the higher temperature (300 °C), the reactions whose migration energy or reaction energy barrier is 1.40 eV can occur within 2.82 s, while the migration or reaction energy barrier of 3.34 eV or higher is still too high for reactions to occur on the time scales of experiments. Note, that there is no reaction whose migration or reaction energy barrier is in the range between 1.40 and 3.34 eV. Using the above criteria, we have analyzed possible reactions between defects summarized in Table 3.

First of all, it is known that FPs are introduced as the direct consequence of radiation so how they are recovered affects the tolerance to radiation-induced amorphization. In the two MAB phases studied here, most vacancies and interstitials in the two MAB phases can be removed through FP recombination. $V_{Al}$ and $Al_I$ in MoAlB can be removed through reaction #7 in Table 3 with the aid of the negative reaction energy, the low energy barrier (barrierless), and the low migration energy of $Al_I$ (0.37 eV). $V_{Al}$ and $Al_I$ in $Fe_2AlB_2$ can also be removed through reaction #7 with the aid of the negative reaction energy, the low energy barrier (barrierless), and the low migration energy of $Al_I$ (0.25 eV). As for $V_B$ and $B_I$, they can be removed through FP recombination (reaction #9) in MoAlB (negative reaction energy, $V_B$ migration energy of 0.68 eV, and energy barrier of 1.33 eV) and in $Fe_2AlB_2$ (negative reaction energy, $V_B$ migration energy of 0.50 eV, and energy barrier of 1.02 eV). In addition, $V_B$ and $B_I$ can recombine through another path, i.e., via formation of $B_{Al}$. In MoAlB, $B_I$ can easily become $B_{Al}$ through reaction #11 ( negative reaction energy, $V_{Al}$ migration energy of 0.46 eV, and barrierless reaction), and then $B_{Al}$ easily reacts with $V_B$ through reaction #3 (negative reaction energy, $V_B$ migration energy of 0.68 eV, and barrierless reaction). In $Fe_2AlB_2$, $B_I$ can become $B_{Al}$ through reaction #11 (negative reaction energy, $B_I$ migration energy of 0.92 eV, and barrierless reaction), and then $B_{Al}$ reacts with $V_B$ through reaction #3 (negative reaction energy, $V_B$ migration energy of 0.50 eV, and low energy barrier of 0.12 eV). As discussed earlier, $Mo_I$ is not stable in MoAlB, and instead it forms $Mo_{Al}$, hence we do not consider the FP recombination of Mo. Meanwhile, $Fe_I$ in $Fe_2AlB_2$ can be removed through reaction #8 with the negative reaction energy, $Fe_I$ migration energy of 1.07 eV, and barrierless reaction.

$Al_B$ in both MAB phases can be removed through reaction #1. In MoAlB, the reaction energy is negative, $V_{Al}$ migration energy is 0.46 eV and the reaction is barrierless, whereas in $Fe_2AlB_2$, the reaction energy is negative, $V_{Al}$ migration energy is 1.17 eV and the reaction is barrierless. $Al_M$ in the two MAB phases can be removed through reaction #2 in MoAlB (negative reaction energy, $V_{Al}$ migration energy of 0.46 eV and low energy barrier of 0.19 eV), and $Fe_2AlB_2$ (negative reaction energy, $V_{Al}$ migration energy of 1.17 eV and barrierless reaction). Next, $B_M$ in both MAB phases can be removed by reaction #4 with the aid of the negative reaction energies, low energy barriers (barrierless for MoAlB and 0.07 eV for $Fe_2AlB_2$) and/or the low migration energies of $V_B$ (0.68 eV for MoAlB and 0.50 eV for $Fe_2AlB_2$). And as mentioned above, $B_{Al}$ can be removed through reaction #3 while providing an intermediate site for the recombination of $V_B$ and $B_I$.

Before looking into $Mo_{Al}$ in MoAlB, we should note that a significant number of antisites $Mo_{Al}$ can form due to the unstable $Mo_I$ as well as from the direct consequence of radiation. However, $Mo_{Al}$ cannot be easily removed in MoAlB even at the temperature of 300 °C because of the high migration energy of $V_{Mo}$ (5.27 eV), which is still too high for the migration to occur. Recall that we assume that the reactions whose migration or energy barrier is higher than 3.34 eV cannot occur at 300 °C. As discussed earlier, although the existence of $Mo_{Al}$ near $V_{Mo}$ can reduce the migration energy to 4.84 eV, it is still too high for the migration to occur. Finally, the only reaction which can remove this defect (reaction #5) cannot occur. There are other potential reactions that could anneal $Mo_{Al}$ (i.e., $B_I + Mo_{Al} \rightarrow B_{Al} + Mo_I$, or $Mo_{Al} + Al_{Mo} \rightarrow Al_{Al} + Mo_{Mo}$). However, the former reaction does not occur due to the formation of an unstable $Mo_I$, and the latter, which is the exchange of antisites, is energetically prohibitive.

In contrast, $Fe_{Al}$ in $Fe_2AlB_2$ can be removed through reaction #12 with the aid of the negative reaction energy, low migration energy of $Al_I$ (0.25 eV), and the barrierless reaction. Although this process creates $Fe_I$, this interstitial can be removed through reaction #8 as discussed earlier.

Lastly, $M_B$ in both MAB phases can be removed through reaction #14 with the aid of the negative reaction energies, low migration energies of MoAlB (0.46 eV) and $Fe_2AlB_2$ (1.17 eV), and low reaction energy barriers (barrierless). This reaction path creates $V_B$ with $M_{Al}$ antisites, whose behavior is distinct in the two MAB phases as discussed earlier, i.e., $Mo_{Al}$ cannot be removed whereas $Fe_{Al}$ can be removed. Therefore, this reaction creates the defect species ($Mo_{Al}$) that cannot be removed in MoAlB, whereas it can contribute to the recovery process in $Fe_2AlB_2$. Changing the energy criterion for the higher temperature (300 °C) does not change the defect behaviors of the two MAB phases. We assumed that at 300 °C the reactions with migration energy or reaction energy lower or equal to 1.6 eV can occur within a few minutes, while the migration or reaction energy barrier of 3.34 eV or higher is too high for the reactions to occur. The migration energy of $V_{Mo}$ (5.27 eV) in reaction #8 for removing $Mo_{Al}$ is still too high in this energy range, whereas all the defects in $Fe_2AlB_2$ can be annealed out.

In summary, the increased tolerance to radiation-induced amorphization of $Fe_2AlB_2$ as compared to MoAlB can be rationalized by the increased production of, and difficulty in annealing out, antisites in MoAlB. In MoAlB, unstable interstitial $Mo_I$ (and therefore unstable Mo FP) is expected to lead to a larger production of $Mo_{Al}$ antisites, which are difficult to anneal out due to the high migration energy of $V_{Mo}$. In addition, there is one defect ($Mo_{Al}$) in MoAlB that is difficult to anneal out even at 300 °C, whereas $Fe_2AlB_2$ has no such defects. In $Fe_2AlB_2$, all the defects are expected to anneal out in a reasonable period of time at both 150 °C and 300 °C.

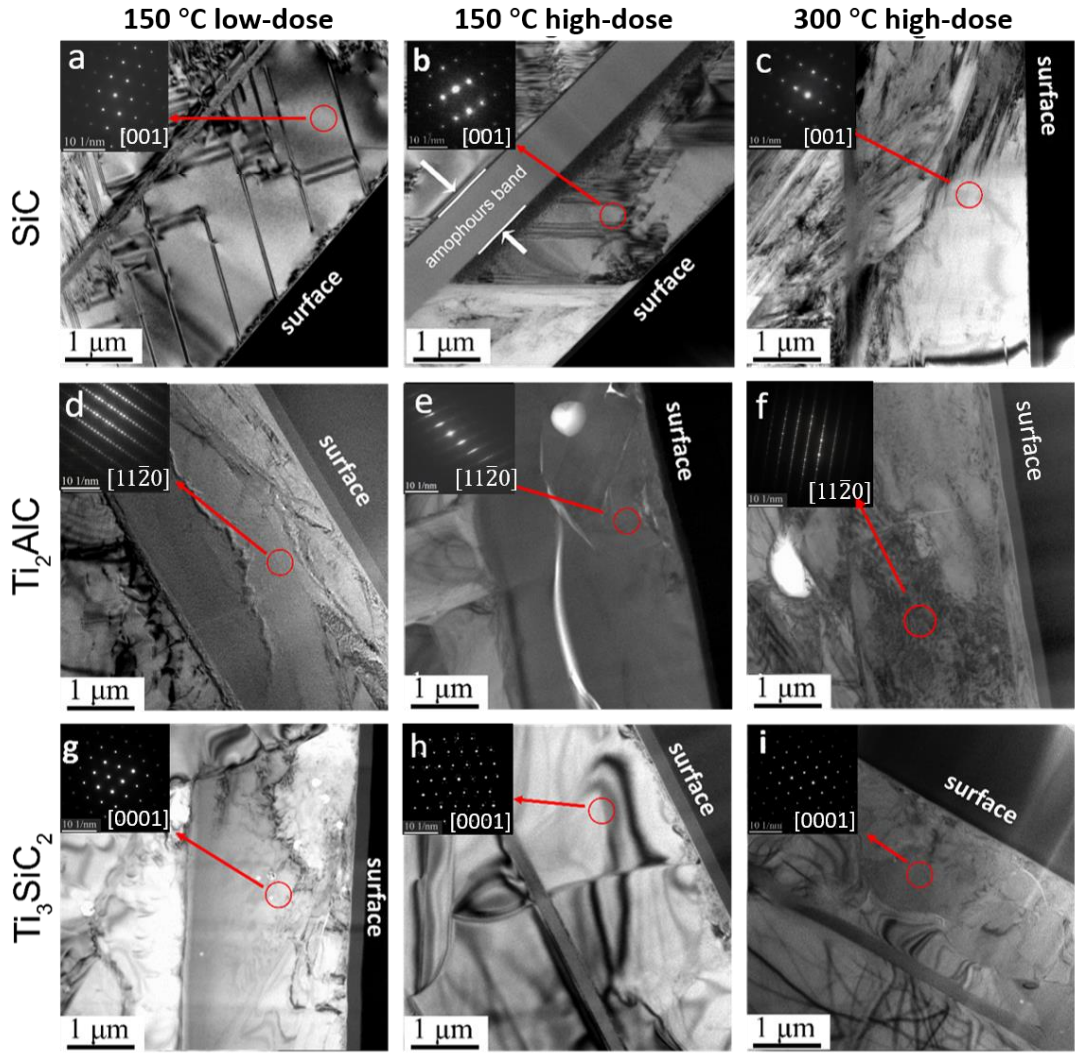

*Figure 8 TEM images of SiC (top row), MAX phases $Ti_2AlC$ (middle row) and $Ti_3SiC_2$ (bottom row), irradiated at $7.5 \times 10^{16}$ ions·cm$^{-2}$ at 150 °C (left column), at $1.5 \times 10^{17}$ ions·cm$^{-2}$ at 150 °C (middle column), and at $1.5 \times 10^{17}$ ions·cm$^{-2}$ at 300 °C (right column). The zone axis of the red-circled region is label in the correlated SAED pattern. The incident beam is perpendicular to the surface of the sample. The light-colored thin band on the top of the surface is the Pt protective layer deposited by electron deposition followed by the thicker, dark colored Pt protective layer deposited by ion deposition during FIB.*

Since $Fe_2AlB_2$ and $MoAlB$ MAB phase materials are both nano-layered ternary borides, which have similar structures to the MAX phase materials, it is instructive to compare radiation resistance of MAB to that of selected MAX phases. We specifically chose $Ti_3SiC_2$ and $Ti_2AlC$, since $Ti_3SiC_2$ is a MAX phase with an unusually high resistance to radiation-induced amorphization and $Ti_2AlC$ contains Al, just like the MAB phases considered here. We are also including comparison to SiC, which is considered to have excellent radiation resistance [42] and is a highly promising material for cladding in nuclear reactor technologies. Cross-sectional TEM images of SiC, $Ti_2AlC$, and $Ti_3SiC_2$ irradiated simultaneously with the MAB phase materials at $7.5 \times 10^{16}$ ions·cm$^{-2}$ (low dose) at 150 °C, $1.5 \times 10^{17}$ ions·cm$^{-2}$ (high dose) at 150 °C, and high dose at 300 °C are shown in Fig. 8.

All the samples irradiated at low dose at 150 °C remained crystalline after irradiation as evidenced by clear diffraction spots in the SAED patterns shown in the insets in Fig. 8. In all samples, irradiation produced many small defect clusters throughout the irradiated regions visible as black-spot defects, but no amorphization in the flat damage region could be found (see Figs. 8a, 8d, and 8g). A very thin (~0.1 µm) amorphous band was formed in the implanted region (at 2.5 µm to 2.6 µm depth) in the CVD SiC as shown in Fig. 8a. There is no obvious amorphous band in $Ti_2AlC$ and $Ti_3SiC_2$, but there are many irradiation-induced cracks on the surface of $Ti_2AlC$ as shown in Fig. 6. The cracks are more than 10 µm long and they extend throughout the entire irradiation range confirmed by TEM. No obvious cracks were found on the surface of $Ti_3SiC_2$ at this dose.

As the irradiation fluence increased to $1.5 \times 10^{17}$ ions·cm$^{-2}$ at 150 °C, there was an obvious amorphous band (more than 0.9 µm wide), formed around the implantation peak in the CVD SiC. According to Fig. 1, the radiation dose in that region is ~ 8 dpa at the edge of this band that is closer to the sample surface, rising up to 30 dpa at the peak region. The region where the dose was relatively flat and equal approximately 1.0 to 2.1 dpa (see Fig. 1) remained crystalline. The density of the black spot defects appears to be very high near the amorphous band region. Under the same irradiation conditions, $Ti_2AlC$ and $Ti_3SiC_2$ still exhibit crystallinity, as evidenced by clear diffraction spots in SAED in the insets of the figures. Some of the diffraction spots disappeared in the SAED pattern of $Ti_2AlC$ (Fig. 8e) which means the damage was significant at this dose. Many irradiation-induced cracks concentrated in the irradiated region were found in the irradiated $Ti_2AlC$, as shown in Fig. 8e. An obvious void as well as a rougher surface can be observed in the irradiated region of Ti$_2$AlC shown in Fig. 8e which should correspond to the bulges on the surface from SEM pictures (Fig. 5). There is no amorphous band at all, even at the peak region. Based on the SRIM results in Fig.1, the radiation dose in the peak region is ~26 dpa for Ti$_2$AlC and ~ 29 dpa for Ti$_3$SiC$_2$. More irradiation induced cracks could be found on the surface of Ti$_2$AlC at 150 °C, high dose irradiation as shown in SEM pictures of Fig. 5. These cracks are larger than those observed in the 150 °C, low dose irradiated sample and tend to connect with each other to form a crack network. There are also some small cracks seen on the surface of Ti$_3$SiC$_2$ at this dose shown in Fig. 5 but this behavior was not observed in the deeper region from TEM results.

For the high dose irradiation at 300 °C, the diffraction patterns of all samples also showed clear diffraction spots, indicating the samples were still crystalline after irradiation. A small amorphous band with a width of only ~0.1 μm could be found in SiC (Fig. 8c) indicating the dose to amorphization increases from ~8 dpa to ~20 dpa at 300 °C by comparing with the 150 °C high dose result. For Ti$_2$AlC, no obvious irradiation-induced cracks were found in the TEM image (Fig. 8f), indicating there were fewer cracks than observed in the 150 °C, high dose irradiated sample. However, many irradiation-induced cracks can still be observed from the SEM pictures as shown in Fig. 5. An obvious void with the size of about 1 μm in diameter can be seen in the near peak region. The surface roughness is also smoother than in the 150 °C, high dose sample. Even though the irradiation dose was relatively high, no obvious phase transformations were found in either Ti$_2$AlC or Ti$_3$SiC$_2$, which have been observed in some parts of ion irradiated MAX phase materials in earlier studies [43–45]. However, partial phase transformation might have occurred in some areas and may require higher magnification TEM or HRTEM to detect.

Our experiments have shown that radiation resistance to amorphization of Fe$_2$AlB$_2$ is comparable to that of SiC. One should note that the resistance of Ti$_2$AlC and Ti$_3$SiC$_2$ to radiation-induced amorphization is better than that of the two MAB phases, but the MAX phases show cracking whereas the MAB phases do not. In addition, Fe$_2$AlB$_2$ has already been shown to have a high decomposition temperature [5] and cracking resistance [15], which is beneficial for nuclear reactor applications.

**Conclusion**

TEM analysis showed that Fe$_2$AlB$_2$ remains fully crystalline under irradiation of $7.5 \times 10^{16}$ ions·cm$^{-2}$ at 150 °C and $1.5 \times 10^{17}$ ions·cm$^{-2}$ at 300 °C, while showing partial amorphization under irradiation of $1.5\times10^{17}$ ions·cm$^{-2}$ at 150 °C. In contrast, MoAlB became amorphous under identical irradiation conditions. On the basis of our first-principle calculations, we were able to rationalize our experimental results. In MoAlB, Mo$_I$ cannot form in the lattice and instead it is expected to create many Mo$_{Al}$ antisites. These antisites cannot be easily removed due to the high migration energy (5.27 eV) of V$_{Mo}$. In contrast, all the defects in Fe$_2$AlB$_2$ are expected to anneal out at both 150 °C and 300 °C. We also performed radiation studies on CVD SiC, MAX phase Ti$_2$AlC, and Ti$_3$SiC$_2$ with the same irradiation conditions as used in the MAB phases. The MAX phases showed that they are tolerant to radiation-induced amorphization under all the irradiation conditions, whereas CVD SiC showed similar trends to Fe$_2$AlB$_2$. Specifically, SiC got amorphized under irradiation of $1.5\times10^{17}$ ions·cm$^{-2}$ at 150 °C and remained crystalline under the other conditions. Our experiments also revealed that the MAX phases showed radiation-induced cracking, which was not found in the MAB phases.

Our study points to MAB phases as a promising class of materials for applications in environments that involve radiation and potentially corrosion (as explained in the introduction of this paper). Numerous MAB phases have been predicted theoretically, and further studies are needed to explore the full potential of these materials for applications in harsh environments.

**Acknowledgment**

The authors gratefully acknowledge financial support from the U.S. Department of Energy, Office of Science, Basic Energy Sciences under Award # DEFG02-08ER46493.

**Supporting information**

1. Determination of the chemical potentials

In order for MoAlB to form, its formation energy is required as below,

$$E_{Total}(MoAlB) = \sum \mu_i^{MoAlB} \quad (i = Mo, Al, B) \qquad \text{Eq. S1}$$

$$E_f(MoAlB) = E_{Total}(MoAlB) - \sum \mu_i^{Bulk}$$
$$= \sum \Delta\mu_i \quad (i = Mo, Al, B) \qquad \text{Eq. S2}$$

where $E_{Total}$ is the total energy of MoAlB calculated, $E_f$ is the formation energy required to form MoAlB, $\mu_i^{MoAlB}$ is the chemical potential of constituent element in MoAlB, and $\mu_i^{Bulk}$ is the chemical potential from its bulk. The calculated $E_f(MoAlB)$ is $-1.36$ eV. Next, to prevent formation of competing binary phases, following conditions are required.

$$E_f(Al_8Mo_3) > 8\Delta\mu_{Al} + 3\Delta\mu_{Mo} \qquad \text{Eq. S3.a}$$
$$E_f(MoB) > \Delta\mu_{Mo} + \Delta\mu_B \qquad \text{Eq. S3.b}$$
$$E_f(Al_{23}B_{50}) > 23\Delta\mu_{Al} + 50\Delta\mu_B \qquad \text{Eq. S3.c}$$

The binaries Al8Mo3 and MoB were selected because they are found in synthesized MoAlB samples[1,4], while Al23B50 is chosen because that is the most stable phase (one with the lowest formation energy) of Al-B system. The calculated formation energies of Al8Mo3, MoB, and Al23Ba are $-3.59$ eV, $-1.04$ eV, and $-4.27$ eV, respectively. Finally, preventing the precipitation of elemental solid from MoAlB requires the following equation.

$$\mu_i^{Bulk} > \mu_i^{MoAlB} \quad (i = Mo, Al, B) \qquad \text{Eq. S4}$$

Using Eq. S1-4 and the calculated formation energies, the chemical potential map of MoAlB was determined and shown in Fig. 6 of the main text. Using the same process, the chemical potential map of Fe2AlB2 was determined, with the calculated formation energies of Fe2AlB2 ($-2.00$ eV) and binaries found in synthesized Fe2AlB2[35]: Al13Fe4 ($-5.67$ eV), AlFe ($-0.66$ eV), and FeB ($-0.76$ eV). The chemical potential set ($\mu_{Mo/Fe}$, $\mu_{Al}$, $\mu_B$) for the characteristic points (A–I, X and Y) are tabulated in Table S1.

***Table S1*** *the chemical potential set ($\mu_{Mo/Fe}$, $\mu_{Al}$, $\mu_B$) for the characteristic points (A–I, X and Y) indicated in Fig. 6 of the main text.*

| eV | MoAlB | | | Fe$_2$AlB$_2$ | | |
|---|---|---|---|---|---|---|
| | $\mu_{Mo}$ | $\mu_{Al}$ | $\mu_B$ | $\mu_{Fe}$ | $\mu_{Al}$ | $\mu_B$ |
| A | −12.20 | −3.75 | −6.79 | −9.15 | −3.75 | −6.79 |
| B | −12.10 | −3.94 | −6.70 | −9.15 | −3.94 | −6.70 |
| C | −11.96 | −4.08 | −6.70 | −9.00 | −4.23 | −6.70 |
| D | −10.92 | −4.08 | −7.74 | −8.24 | −4.23 | −7.47 |
| E | −10.92 | −4.20 | −7.62 | −8.24 | −4.19 | −7.49 |
| F | −12.12 | −3.75 | −6.87 | −9.16 | −3.91 | −6.70 |
| G | −11.26 | −4.06 | −7.41 | −8.43 | −4.23 | −7.27 |
| H | | | | −8.90 | −3.75 | −7.04 |
| I | | | | −8.24 | −4.41 | −7.37 |
| X | −11.48 | −3.99 | −7.27 | −8.64 | −3.99 | −7.18 |
| Y | −11.80 | −3.87 | −7.07 | −9.13 | −3.87 | −6.74 |

2. The most stable interstitial sites for MoAlB and Fe$_2$AlB$_2$

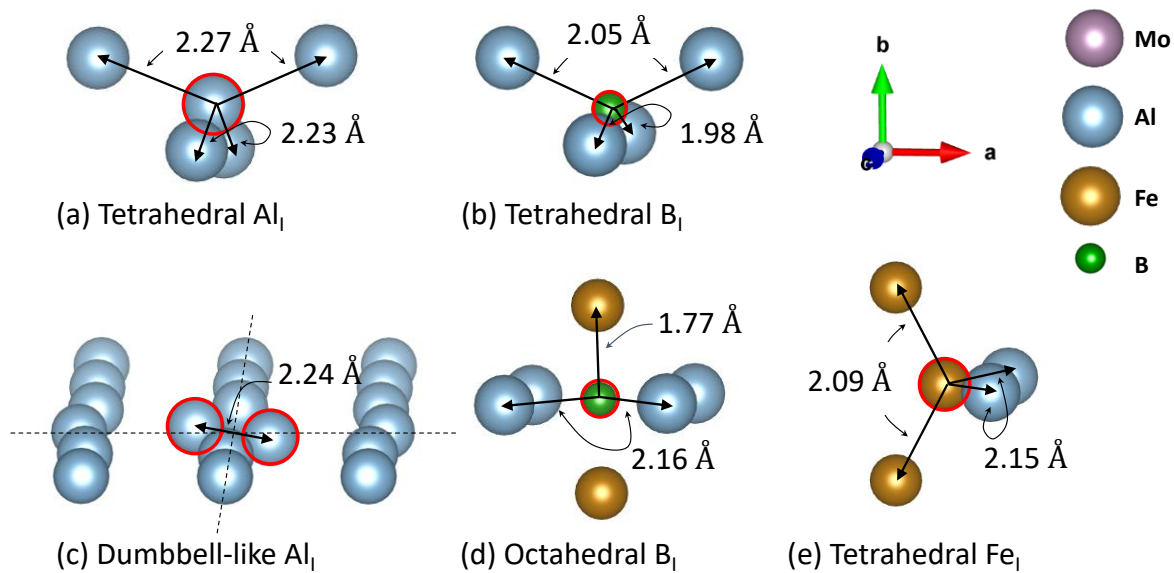

*Figure S2* *Schematics of the most stable interstitial sites for MoAlB and Fe$_2$AlB$_2$ with the interatomic distances shown. (a) and (b) are Al$_I$ and B$_I$ in MoAlB, respectively. (c), (d), and (e) are Al$_I$, B$_I$, and Fe$_I$, respectively, in Fe$_2$AlB$_2$. Note that Mo$_I$ does not exist because Mo$_I$ is unstable and forms Mo$_{Al}$ and Al$_I$ (see Table 1 of the main text).*

| Dose | $a$/Å | $b$/Å | $c$/Å |
| --- | --- | --- | --- |
| Unirradiated | 2.913 | 11.003 | 2.861 |
| 150 °C -low dose | 2.918 | 10.969 | 2.853 |
| 300 °C-high dose | 2.933 | 10.994 | 2.860 |
| 150 °C -high dose | 2.949 | 10.903 | 2.836 |

*Tabel S1: lattice parameter changes before irradiation and after irradiation at $7.5 \times 10^{16}$ ions·cm$^{-2}$ at 150 °C, at $1.5 \times 10^{17}$ ions·cm$^{-2}$ at 150 °C, and at $1.5 \times 10^{17}$ ions·cm$^{-2}$ at 300 °C*